\documentclass[sigconf]{acmart}
\AtBeginDocument{%
  \providecommand\BibTeX{{%
    \normalfont B\kern-0.5em{\scshape i\kern-0.25em b}\kern-0.8em\TeX}}}

\copyrightyear{2020}
\acmYear{2020}
\acmDOI{10.1145/3396851.3397692}

\acmConference[e-Energy'20]{The Eleventh ACM International Conference on Future Energy Systems}{June 22--26, 2020}{Virtual Event, Australia}
\acmBooktitle{The Eleventh ACM International Conference on Future Energy Systems (e-Energy'20), June 22--26, 2020, Virtual Event, Australia}
\usepackage{subfigure}
\usepackage{algorithmic} 
\usepackage{amssymb}
\usepackage{booktabs}
\begin{document}

\title{Interaction Between Coordinated and Droop Control PV Inverters}
% \author{Peter Lusis, Lachlan Andrew, Ariel Liebman, Guido Tack}

\author{Peter Lusis}
\email{peter.lusis@monash.edu}
\affiliation{%
  \institution{Monash University}
  \city{Melbourne}
  \country{Australia}
}
\author{Lachlan L. H. Andrew}
\email{Lachlan.Andrew@unimelb.edu}
\affiliation{%
  \institution{The University of Melbourne}
  \city{Melbourne}
  \country{Australia}
}
\author{Ariel Liebman}
\email{ariel.liebman@monash.edu}
\affiliation{%
  \institution{Monash University}
  \city{Melbourne}
  \country{Australia}
}
\author{Guido Tack}
\email{guido.tack@monash.edu}
\affiliation{%
  \institution{Monash University}
  \city{Melbourne}
  \country{Australia}
}
\orcid{0000-0003-3357-6498}

% \author{PL}

\def\la#1{\textbf{$<<$#1$>>$}}
\def\pl#1{\textcolor{red}{$<<$#1$>>$}}
\renewcommand{\shortauthors}{P. Lusis, L. L. H. Andrew,  A. Liebman and G. Tack}

\begin{abstract}
Autonomous droop control PV inverters have improved voltage regulation compared to the inverters without grid support functions, but more flexible control techniques will be required as the number of solar photovoltaic (PV) installations increases. This paper studies three inverter future deployment scenarios with droop control inverters, non-exporting inverters, and coordinated inverter control (CIC). The network operation and the interaction between various inverter control methods are studied by simulating inverter operation on two low-voltage networks. Considering 30\% PV penetration as the base case, we demonstrate that coordinated inverters can mitigate overvoltages and voltage fluctuations caused by the tripping of passive inverters in 85\% of PV location cases when at least as many coordinated as passive inverters are deployed on the 114-node test feeder. However, this rate reduced to 37\% with the IEEE 906-node network demonstrating that the deployment of coordinated inverter control may not be able to reverse passive inverter-related voltage disturbances when the build-up of passive inverters has reached a certain threshold.  

The aggregated PV output from coordinated inverters can be also used to provide grid support services. When the low-voltage networks operate close to the upper voltage limits, the change in the power output from coordinated inverters following a regulation request may be partially offset by passive inverters. Considering an equal number of passive and coordinated inverters, this paper shows that for each unit of the down-regulation request delivered by coordinated inverters, passive inverter output may increase by up to 0.2 units and decrease by up to 0.45 units during coordinated inverter up-regulation. 
 
\end{abstract}



%%
%% Keywords. The author(s) should pick words that accurately describe%% the work being presented. Separate the keywords with commas.

\maketitle

% ======================================== INTRODUCTION ==================
\section{Introduction}\label{Introduction}
Environmental awareness, supportive policy mechanisms and falling costs of solar photovoltaic (PV) components are driving an accelerating uptake of distributed energy resources in low-voltage (LV) residential areas. It has become a challenge for many distribution network service providers (DNSPs) to host new solar PV installations~\cite{Chaudhary2018}. Additional PV output generated in the same area results in voltage surges, thus increasing the risk of appliance damage. The consequent overvoltage protection settings will disconnect PV inverters and can also initiate voltage fluctuations across the network~\cite{Farivar2011}.The voltages issues are exacerbated in the networks with a high build-up of, so-called, \emph{legacy} inverters without grid support functions~\cite{Wyndham2016}. 

% === PASSIVE CONTROL ===
Multiple studies have shown improved voltage control using autonomous inverters with reactive power (Volt/VAr) droop curve~\cite{Seuss2016, Anonas2018, Ali2018}, active power (Volt/Watt) droop curve~\cite{GhapandarKashani2017, Kraiczy2018, Ali2015}, and a combination of both~\cite{Giraldez2017,Giraldez2018}. All new solar PV installations must include autonomous voltage regulation based on a droop control strategy as required under the IEEE 1547 Standard on Smart Inverters~\cite{Stice2019}. Autonomous inverters are network-agnostic, which makes them simple and easy to deploy, but it also means that they provide only suboptimal operations of the network~\cite{Baker2018}, particularly in the areas with a large number of customers per distribution transformer~\cite{Parajeles2017}. While passive inverter control can reduce the number of customers experiencing overvoltages~\cite{Hoke2018}, the limitations of such control approach become particularly obvious when operating in a network with high PV penetration~\cite{Lusis2019}. 

% === COORDINATED CONTROL AND HOW THIS PAPER IS DIFFERENT ===
Improved voltage regulation in a PV-rich distribution network was achieved in~\cite{Dallranese2015} by implementing bidirectional communication between  inverters and a central node. It utilises smart meter data of customers' voltages and net loads across the network to determine and deploy the optimal active and reactive power setpoints to each of the controllable inverters. Existing studies have demonstrated the operation of coordinated inverters alone~\cite{Guggilam2016}, in combination with battery storage~\cite{DallAnese2018} and electric vehicles~\cite{Chapman2018}. To the authors' knowledge, all previous studies assume that all distributed energy resources (DER) in the network can be controlled. This is optimistic, given the wide deployment of passive inverters. Moreover, some customers may be unwilling to share inverter data or give up the control over their assets~\cite{Zeraati2018}. Thus, in this paper, we simulate the deployment of coordinated inverters in a LV network with a significant share of legacy and autonomous inverters and demonstrate the aggregated ability of coordinated inverters to mitigate voltage disturbances.

% === COORDINATED CONTROL FOR GRID SUPPORT SERVICES ===
Centralised PV inverter control can also enable grid support services (GSS) and establish new relationships between the system operator and the entity in charge of DER~\cite{Rasmussen2018}. The aggregated output capacity of coordinated inverters can be seen as a \emph{virtual power plant} (VPP) and the entity in charge of coordinating inverters as a \emph{VPP aggregator}~\cite{Pasetti2018}. PV inverters have the flexibility and fast response needed to provide up-regulation ($\mathit{UR}$) and down-regulation ($\mathit{DR}$) services. That means that it can meet the need to change the frequency or respond to an event by increasing/decreasing the power output. The calculation of feasible and flexible operating regions and maximum grid support capacity for a VPP was demonstrated in~\cite{Riaz2019}. However, the network operator may not see the expected change in the frequency following a response from the VPP: since ($\mathit{UR}$) and ($\mathit{DR}$) delivery by coordinated inverters alter local voltages, it will trigger inverters with passive control to change their output (as per local voltage control settings). Autonomous inverters' sliding along the droop curve or potential disconnection due to overvoltages induced by the VPP aggregator should be quantified and accounted for before making a GSS capacity offer to the market operator. 

By simulating increasing PV penetration levels, we demonstrate that coordinated inverters are significantly more effective in voltage regulation than autonomous inverters while avoiding unnecessary curtailment, which occurs in the scenario with non-exporting inverters. We have also demonstrated that offering grid support services can serve as a financial incentive for the CIC deployment. This paper can be distinguished from other studies by the fact that we have considered the existing inverter infrastructure in LV distribution networks, with 30\% PV penetration level consisting of existing autonomous and legacy inverters. The cumulative effects of inverter interaction are quantified in terms of voltage disturbances and power curtailment. For the given networks, it is shown that, by deploying as many coordinated inverters as the existing passive inverters, voltage issues can be prevented in 85\% of PV locations.

% === THE CONTENT OF THE PAPER
This paper is organised as follows: Inverter models are introduced in Section 2, and the network operation and simulation setup are explained in Section 3. The provision of grid support services by coordinated inverters is formulated in Section 4. The results are discussed in Section 5 and a summary of the results and conclusions is given in Section 6. 
\section{Inverter Models}\label{Methodology}
% ===================================================
Before introducing the models of the four types of solar PV inverter studied in
this paper, it is useful to define notation for the network.
The network under study consists of a set $\mathcal{N}$ of $N$ nodes that have a
load and, in some cases, a PV inverter. We often consider an extended network
$\mathcal{N'} = \mathcal{N} \cup \{\mathbf{s}\}$, where $\mathbf{s}$ denotes
the secondary side of the distribution transformer (the slack bus).
Lines are represented as $(m,n) \in \mathcal{M} \subset \mathcal{N'} \times \mathcal{N'}$.

In our studies, the network initially hosts $L$ legacy and $A$ autonomous inverters (described below) located on the nodes of the sets $\mathcal{L}$ and $\mathcal{A}$, respectively. We increase PV penetration by adding new solar PV systems equipped with one of the three following inverter models: coordinated inverters, non-exporting inverters or more autonomous inverters. Subscripts $l\in\mathcal{L}, a\in\mathcal{A}, c\in\mathcal{C}$, and $e\in\mathcal{E}$ denote the quantities pertaining to each legacy inverter $l$, autonomous inverter $a$, coordinated inverter $c$, and non-exporting inverter $e$. All inverters have the same rated apparent power capacity $\textbf{S}$, which sets the limits on the maximum active power injection $P^\text{inj}$ and reactive power support $Q$.
Complex voltages are expressed per unit and denoted by $V$ with $\Re$ and $\Im$ as the real and imaginary components. Constants beyond our control are written in bold. The individual inverter models are as follows.
% ========================= Legacy =============================
\subsection{Legacy Inverters}\label{sec:LEG}
% ========================= Legacy =============================
Legacy inverters $l\in\mathcal{L}$ are not equipped with grid support functions and can operate only at unity power factor. The operating state of legacy inverters is fully captured by a binary variable $u_l$, where $u_l=1$ if the inverter is operational, and $u_l=0$ if it has tripped due to overvoltage; this is referred to as the status of the inverter. The injected active power $P^\text{inj}_l$ is 
 \begin{equation}
   P^\text{inj}_l=\begin{cases}
     \textbf{P}^\text{av}_l, & \text{if } u_l=1,\quad \forall l \in \mathcal{L},\\
     0, & \text{otherwise}.
   \end{cases}
 \label{eq:LegacyInv}
 \end{equation} 
% ========================= Autonomous =============================
\subsection{Autonomous Inverters}\label{sec:AUT}
\begin{figure}[t] 
\centerline{\includegraphics[width= \linewidth]{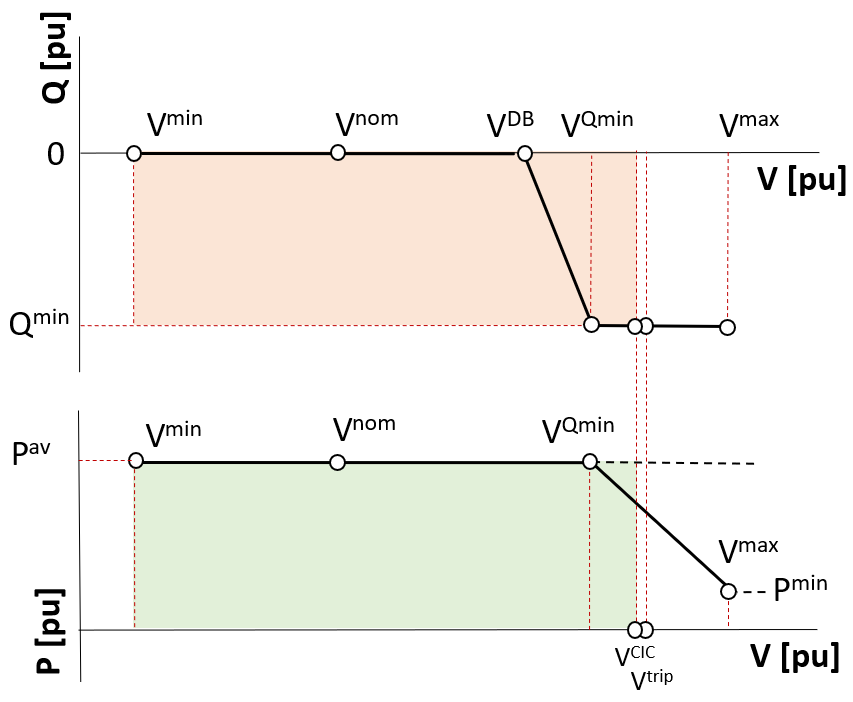}}
\caption{Volt/VAr and Volt/Watt droop curves on a per-unit basis. The highlighted areas represent the feasible setpoint space for coordinated inverter control.} 
\label{fig:droopcruves} 
\end{figure}

% ========================= Autonomous =============================
The operating state and output of non-coordinated inverters are dependent only on the local voltage measured by the inverter. An autonomous inverter $a\in\mathcal{A}$ enables voltage regulation using a combination of Volt/VAr and Volt/Watt droop curves (Fig.~\ref{fig:droopcruves}). The feasible region of the full Volt/VAr droop curve is a piecewise linear function with three constant sections and two decreasing ones bounded by voltage setpoints. However, the Volt/VAr function considered here consists of three pieces since we neglect the control of undervoltage for the scenarios studied in this paper. Given the measured inverter output voltage $V_a$, the nominal inverter reactive power injection in the grid operating in Volt/VAr response mode is 
\begin{equation}
    Q^{pu}_a =  \textbf{m}^\text{V}V_a+\textbf{c}^\text{V}, \quad \forall a \in \mathcal{A}.
    \label{eq:Q_VV}
\end{equation}
The slope $\textbf{m}^\text{V}$ and intercept $\textbf{c}^\text{V}$ of the Volt/VAr droop curve are 
\begin{align} 
    \textbf{m}^\text{V}=- \textbf{Q}^{\min,pu}/({\textbf{V}^\text{Qmin} - \textbf{V}^\text{DB}}),
    \label{eq:slopeVV} 
    \\
    \textbf{c}^\text{V}=\textbf{Q}^{\min,pu}/\textbf{m}{\textbf{V}^\text{Qmin}}, 
    \label{eq:intercVV} 
\end{align}
where $\textbf{V}^\text{DB}$ is the upper reference point of the deadband, and $\textbf{V}^\text{Qmin}$ denotes the voltage level at which the maximum reactive power support is reached. 

Recognising that inverters may use reactive power to regulate voltage without compromising the active power injection, autonomous inverters are operated in reactive power priority mode. That means that voltage regulation is first attempted through reactive power support. The Volt/Watt response mode is activated when inverter voltage cannot be maintained within operational limits using reactive power. 
The Volt/Watt function linearly reduces active power output until the cut-off voltage $\textbf{V}^{\max}$ is reached, as
\begin{equation}
    P^{pu}_a= \textbf{m}^\text{W}V_a+\textbf{c}^\text{W},\quad \forall a \in \mathcal{A}.
    \label{eq:Q_VW}
\end{equation}
The slope $\textbf{m}^\text{W}$ and intercept $\textbf{c}^\text{W}$ are derived from the line going through the points $(\textbf{V}^{\textbf{Q}_{\min}}, \textbf{P}^\text{av})$ and $(\textbf{V}^{\max}, \textbf{P}^{\min})$. When operating in Volt/Watt response mode, $Q^{pu}_a = \textbf{Q}^{\min,pu}$. Converting to base units, this gives
\begin{equation}
    Q_a = \textbf{S}Q^{pu}_au_a,\quad \forall a \in \mathcal{A},
    \label{eq:Q_AUT}
\end{equation}
where $u_a\in\{0,1\}$ is the inverter ON/OFF state variable. Then, the inverter injected power $P^\text{inj}_a$ operating in Volt/Watt response mode is 
\begin{equation}
  P^\text{inj}_a = \min\left\{\sqrt{\textbf{S}^2-Q_a^2},\textbf{P}^\text{av}_aP^{pu}_a\right\}u_a, \quad \forall a \in \mathcal{A}.
 \label{eq:Pinj_AUT}
\end{equation}
% ===================================================================
\subsection{Coordinated Inverters}\label{sec:CIC}
% ===================================================================
The optimum active and reactive power target values ($P^\text{inj}_c, Q_c$) for a set of coordinated inverters $c\in\mathcal{C}$ are found by solving the optimisation problem \eqref{eq:Objmain}, which will be introduced in later stages. First, it is subject to a capacity limit
\begin{equation}
(P_c^\text{inj})^2 + Q_c^2 \leq \textbf{S}^2, \quad \forall c \in \mathcal{C}.
\label{eq:CICconst1}
\end{equation}
The injected PV output $P^\text{inj}_c$ from each coordinated inverter is non-negative and limited by the available active power on the secondary side of the PV inverter $\textbf{P}^\text{av}_c$ as
 \begin{equation}
\textbf{0} \leq P^\text{inj}_c \leq \textbf{P}^\text{av}_c, \quad \forall c \in \mathcal{C}.
\label{eq:CICconst2}
\end{equation}
Since we are considering the periods with excess PV output leading to overvoltages, the inverter capability to provide reactive power is ignored. Thus, the maximum reactive power support is limited to 
 \begin{equation}
-\textbf{Q}^{\min,pu}\textbf{S} \leq Q_c \leq \textbf{0}, \quad \forall c \in \mathcal{C}, % \textbf{Q}^{\min} 
\label{eq:CICconst3}
\end{equation}
as in \cite{Hall2019}, where $\textbf{Q}^{\min,pu}$ is the maximum reactive power that the inverter can absorb from the grid.
% ========================= Non-exporting ==========================
\subsection{Non-Exporting Inverters}\label{sec:NonExp}
% ========================= Non-exporting ==========================
The last inverter type considered in this study requires inverters $e\in\mathcal{E}$ to be connected in a ``self-supply'' mode. This non-exporting setup prevents reverse power flow at the point of common coupling of the customer premises. The non-exporting mode can be seen as the most neutral future direction for distribution networks since the inverters neither contribute to the voltage rise nor provide grid support functions. It is also arguably the fairest model to the customers as all PV systems are equally affected. The excess PV output has to be either stored locally for later or curtailed. For this study, we only consider curtailment, and leave the evaluation of energy storage for future work. The injected power from non-exporting inverters $P^\text{inj}_e$ is 
\begin{equation}
  P^\text{inj}_e=\begin{cases}
  \min\left\{\textbf{P}^\text{av}_{e},\textbf{P}^\text{d}_{e}\right\}, & \text{if } u_e=1,\quad \forall e \in \mathcal{E}, \\
  0, & \text{otherwise},
  \end{cases}
  \label{eq:Non-Exp}
\end{equation} 
where $u_{e}\in\{0,1\}$ denotes the status of the inverter.
 
\section{Inverter Control Scheme}\label{Simulations}
% ====================================================================
We assume that the net active and reactive power demand ($\textbf{P}^{\text{net}}_{n,t}$, $\textbf{Q}^{\text{net}}_{n,t}$) and the available PV output data $\textbf{P}^{\text{av}}_{c,t}$ are available at each time $t$. In practice, this data is obtained from the Advanced Metering Infrastructure (AMI) and the coordinated inverters\footnote{When no power is being curtailed, this is available from the maximum power point tracker. When power is being curtailed, this could be measured by performing occasional maximum power point probing.}. While the status and PV output from legacy and autonomous inverters are unknown to the AMI, we will show that the information about the net demand is sufficient to compute the optimum active and reactive power target values ($P^\text{inj}_c, Q_c$), which are to be deployed on all coordinated inverters.

During the periods of peak PV production, voltage disturbances are often related to legacy and autonomous inverters shutting off. Under Australian standards, this happens if the rolling 10-minute average voltage exceeds the tripping voltage $\textbf{V}^\text{trip}$. Coordinated inverter control (CIC) seeks to meet a more stringent target of preventing any node from exceeding this threshold. Thus, we define $\textbf{V}^\text{CIC}$, which is set at 0.5\% of $\textbf{V}^\text{nom}$ lower than $\textbf{V}^\text{trip}$ to provide a reserve to the tripping voltage and prevent immediate disconnection of passive inverters. The $\textbf{V}^\text{CIC}$ safeguard against inverter tripping can be made more conservative to account for cases when poor internet connection delays AMI data transfer or CIC setpoints are suboptimal due to the limited information about non-coordinated network nodes. 
% ========================= Autonomous Inverter Control ==========================
\subsection{Passive Inverter Control}\label{PassiveInverter}
% ========================= Autonomous Inverter Control ==========================
The set of passive inverters comprises autonomous and legacy inverters located on the nodes $p \in \mathcal{P} =  \mathcal{A}\bigcup \mathcal{L}$. The operating state and output from passive inverters are dependent only on the local voltage measured by the inverter. Both inverter types share the same protection rules to disconnect from the grid. The governing rule enforces instantaneous disconnection of all passive inverters exceeding $\textbf{V}^{\max}$. Passive inverters will also shut off if the moving average voltage for $\textbf{u}^{\uparrow}$ time intervals is above the \textit{average trip} voltage $\textbf{V}^\text{trip}$. While the average trip voltage condition is not satisfied, legacy inverters will operate at unity power factor, while autonomous inverters will follow droop curve settings.

Spurious voltage oscillations occur in simulations that adjust all inverter setpoints at the same time.  These are mitigated by applying a low-pass filter
\begin{equation}
Q^{\text{pu}}_{a,t} = \left(1-\frac{\Delta T}{\tau^\text{V}}\right) Q^{\text{pu}}_{a,t-1} + \left(\frac{\Delta T}{\tau^\text{V}}\right)\hat Q^{\text{pu}}_{a,t},
\label{eq:tauVV}
\end{equation} 
with a time constant $\tau^{V}$, as demonstrated in \cite{Jahangiri2013}. Alternatively, the voltage oscillations can also be mitigated by limiting the maximum active and reactive power change of autonomous inverters between consecutive time steps, as shown in \cite{Procopiou2020}. Similarly, the active power injection per unit is given by \eqref{eq:Q_VW} with the filter 
\begin{align}
P^{\text{pu}}_{a,t} = \left(1-\frac{\Delta T}{\tau^\text{W}}\right) P^{\text{pu}}_{a,t-1} + \left(\frac{\Delta T}{\tau^\text{W}}\right)\hat P^{\text{pu}}_{a,t}.
\label{eq:tauVW}
\end{align} 

In our simulations, which use discrete time steps of 1 minute, the change in demand and available PV output between consecutive periods can lead to multiple inverters exceeding $\textbf{V}^\text{trip}$ and shutting off. An artefact of the simulation is that this change of state would happen simultaneously. To avoid this artefact, we allow only one passive inverter $p$ to disconnect in each period based on random weighted sampling 
\begin{equation}
w^{\uparrow}_{p,t} =  (\textbf{V}^{\max}-V_{p,t})^{-2}.
\label{eq:weights_Disc}
\end{equation}
Note that voltage magnitude is expressed per unit and $V_{p,t} \in [\textbf{V}^\text{trip}, \textbf{V}^{\max}]$, where $\textbf{V}^{\max}$ is the maximum permissible voltage in the network beyond which autonomous inverters instantly disconnect. Without this constraint, the simulations amplify voltage oscillations beyond what is expected in a real network.  
%The reason for limiting the number of inverters tripping in a single period is to prevent the amplifying effect of voltage oscillations as our time step is 30s that does not represent network dynamics or transient changes. 

Inverters remain disconnected for at least $\textbf{u}^{\downarrow}$ periods and thereafter will reconnect once the voltage recovers to within the continuous operating level $\textbf{V}^\text{trip}$. Again, only one simulated inverter can reconnect at a time based on weighted random sampling with weights
\begin{equation}
w^{\downarrow}_{p,t} =  (V_{p,t}-\textbf{V}^\text{nom})^{-2}.
\label{eq:weights_Rec}
\end{equation} 
During voltage disturbances or faults, non-exporting inverters act in a similar way to legacy inverters, automatically disconnecting from the grid if $\textbf{V}^\text{trip}$ or $\textbf{V}^{\max}$ conditions are met. They remain disconnected until the voltage has stabilised below the $\textbf{V}^\text{trip}$ threshold. While \eqref{eq:weights_Disc} and \eqref{eq:weights_Rec} control the maximum active and reactive power change between consecutive timesteps, they can be overridden by disconnection due to exceeding $\textbf{V}^\text{trip}$ or $\textbf{V}^{\max}$ thresholds. 
% ===================================================================
\subsection{Coordinated Inverter Control}\label{OID}
% ===================================================================
During the periods when the network operates within acceptable voltage levels, coordinated inverter control attempts to minimise active power curtailment 
\begin{equation} 
\phi(P_{\mathcal{C},t})= \sum_{c\in\mathcal{C}} P_{c,t}
\label{eq:CIC_Obj1} 
\end{equation} 
and line active power losses 
\begin{align}
\rho(V_t) = \sum_{(m,n)\in\mathcal{M}}&\Re\{\textbf{y}^*_{mn}\}\Big((\Re\{V_{m,t}\}+\Re\{V_{n,t}\})^2 \nonumber \\
& +(\Im\{V_{m,t}\}+\Im\{V_{n,t}\})^2\Big), 
\label{eq:CIC_Obj2} 
\end{align}
where $\textbf{y}^*_{mn}$ denotes the complex conjugate of admittance between nodes $m$ and $n$. 
The voltage regulation is enforced using a penalty term $ \kappa(V_{d,t})$ with large $\textbf{M}$ for voltages above $\textbf{V}^\text{CIC}$ on the nodes $d\in\mathcal{D}=\mathcal{C}\bigcup\mathcal{P}$ comprising PV systems such that
\begin{align} 
  \kappa(V_{d}) = \sum_{d \in \mathcal{D}} \begin{cases}
     \textbf{M}V_{d,t}, & \text{if } |V_{d,t}|^2>(\textbf{V}^\text{CIC})^2 \\
     0, & \text{otherwise},
  \end{cases}
 \label{eq:CIC_Obj3}
\end{align}
where %the absolute voltage is bounded by $V_d^{abs}\in[0,+\infty)$. 
\begin{equation}
   |V_{d,t}|^2 = (\Re\{V_{d,t}\})^2+(\Im\{V_{d,t}\})^2,\quad \forall d\in\mathcal{D}.
   \label{eq:CIC_QuadConst}
\end{equation}
To improve the algorithm efficiency, We replace \eqref{eq:CIC_QuadConst} by introducing auxiliary variables $x$ and $y$ for piecewise linear approximation of the quadratic terms $(\Re\{V_{d,t}\})^2$ and $(\Im\{V_{d,t}\})^2$ as 
\begin{align}
  x_{d,t} = \begin{cases} \textbf{m}_1\Re\{V_{d,t}\}+\textbf{c}_1 & \text{\ if } \textbf{V}^{\min}\leq\Re\{V_{d,t}\}<\textbf{V}^\text{Qmin},\;d\in\mathcal{D}\} \\
  \textbf{m}_2\Re\{V_{d,t}\}+\textbf{c}_2 & \text{\ if } \textbf{V}^\text{Qmin}\leq\Re\{V_{d,t}\}<\textbf{V}^{\max},\;d\in\mathcal{D}\}
  \\
0, & \text{otherwise},
\end{cases}
   \label{eq:CIC_xConst}
\end{align}
\begin{align}
  y_{d,t} = \begin{cases}  \textbf{m}_3\Im\{V_{d,t}\}+\textbf{c}_3 & \text{\qquad if } 0\leq\Im\{V_{d,t}\}< 0.1,\;d\in\mathcal{D}\} 
  \\
  \textbf{m}_4\Im\{V_{d,t}\}+\textbf{c}_4 & \text{\qquad if } 0.1\leq\Im\{V_{d,t}\}<0.2,\;d\in\mathcal{D}\}
  \\
0, & \text{otherwise}.
\end{cases}
   \label{eq:CIC_yConst}
\end{align}
where parameters $\textbf{V}$ are given in the Appendix. The range of $\Re\{V_d\}$ and $\Im\{V_d\}$ is limited by their min and max feasible values accordingly. To avoid using a multivariate piecewise objective term, we add another auxiliary variable $z_{d,t}$ so that
\begin{align}
   z_{d,t} = x_{d,t} + y_{d,t},\quad \forall d\in\mathcal{D}.
   \label{eq:CIC_zConst}
\end{align}

In addition, it is desirable that households with coordinated PV inverters are self-sufficient when the available PV output $\textbf{P}^{\text{av}}_c$ exceeds local demand $\textbf{P}^d_c$. Thus, we reduce the voltage penalty by $\sqrt{\textbf{M}}$ so that curtailment from coordinated inverters $P_c$ does not affect the household's ability to satisfy its own electricity consumption, by adding an objective term
\begin{align}
    \quad \nu(P_{\mathcal{C},t}) = -\sum_{c \in \mathcal{C}} \begin{cases}
    \sqrt{\textbf{M}} P_{c,t}, & \text{if }  P_{c,t} < \textbf{P}^\text{av}_{c,t}-\textbf{P}^\text{d}_{c,t} \\
     0, & \text{otherwise}.
\end{cases}
\label{eq:CIC_Obj4}
\end{align}

We assume that line impedances and network topology are known when solving the optimisation. The real and imaginary parts of the inverse of the admittance matrix are $\textbf{R}_{mn}$ and $\textbf{X}_{mn}$. Combining this information with the AMI data allows us to calculate the nodal voltage balance such as 
\begin{align}
\Re\{V_{n,t}\} = &\textbf{V}^\text{nom}+\sum_{m\in\mathcal{\mathcal{N}}}\Big(\textbf{X}_{mn}\textbf{Q}^\text{net}_{n,t}  
+ \textbf{R}_{mn}\textbf{P}^\text{net}_{n,t}\Big) + 
\nonumber \\ &
\sum_{c\in\mathcal{C}}\Big( \textbf{R}_{cn}P^\text{inj}_{c,t} + \textbf{X}_{cn}Q_{c,t}\Big), \quad \forall n\in \mathcal{N},
\label{eq:NodalConst1}
\end{align}
\begin{align}
\Im\{V_{n,t}\} = & \sum_{m\in\mathcal{N}} \Big(\textbf{R}_{mn}\textbf{Q}^\text{net}_{n,t} + \textbf{X}_{mn}\textbf{P}^\text{net}_{n,t}\Big) + 
\nonumber \\&
\sum_{c\in\mathcal{C}}\Big( \textbf{X}_{cn}P^\text{inj}_{c,t} -  \textbf{R}_{cn}Q_{c,t}\Big), \quad \forall n\in \mathcal{N}.
\label{eq:NodalConst2} 
\end{align}
This linearisation was introduced in \cite{Dhople2016} and offers fast convergence, as demonstrated in \cite{Guggilam2016}. $\textbf{V}^\text{nom}$ is set to 1.0\,per unit.

Finally, the coordinated inverter control (CIC) is  
\begin{align}
 \underset{V,P_{\text{c}}}{\text{minimise \textit{}}} 
& \phi(P_{\mathcal{C},t}) + \rho(V_t) + \kappa(V_{t}) + \nu(P_{\mathcal{C},t})
\label{eq:Objmain} \\
\text{subject to } & \eqref{eq:CICconst1}-\eqref{eq:CICconst3},\eqref{eq:CIC_xConst}-\eqref{eq:CIC_zConst},\eqref{eq:NodalConst1}-\eqref{eq:NodalConst2},  \nonumber
\end{align} 
taking the form of a quadratically constrained quadratic program (QCQP). 

\section{Case Study}\label{CaseStudy}
%   
% ===================  NETWORK  ========================   
%  
\subsection{Network Topology} 
In the following study, we test inverter control models on two low-voltage distribution networks. The 114-node LV circuit, introduced in \cite{Prettico2016}, is a reference network model developed for European distribution network studies. It is a three-phase balanced LV circuit connected to a 20kV/400V 400kVA secondary transformer. The network is built out of 50mm$^2$ conductors with an $R/X$ ratio of 6:1. The load data and network topology used in this paper are available on request.

The 30-min AMI data for 30 households in the Greater Sydney Area was acquired from \cite{Ausgrid2012}. Data from the $i$th household, $i = 1...30$ was allocated to the nodes with numbers $n\equiv i \mod 30$. Only the daylight hours between 8\;am and 7.30\;pm were considered since no solar-induced voltage issues were recorded outside those hours. Given active power demand, we assume a constant power factor of 0.95 leading. The mean load in the summer period is 0.77\;kW with a standard deviation of 0.27\;kW, while the same values for the winter period are 0.83\;kW and 0.53\;kW, respectively. For the simulations, the household load was up-sampled to 30-second intervals using cubic splines.  

The second network is a modified IEEE 906-bus European test feeder with 55 loads \cite{Espinosa2015}; recommended for LV network studies by the IEEE PES Distribution Systems Analysis Subcommittee \cite{Schneider2017}. Uniform scaling was applied increasing the feeder length by a factor of 2.5 to represent low-density residential areas common in Australia.

Both networks comprise 6\;kWp (kilowatt peak) solar PV systems connected to a 6\;kVA inverter with a reactive power limit of 0.44 of the inverter kVA capacity. The PV system power losses are assumed constant at 17\%, giving a maximum inverter AC output of 5\;kW. Inverter setpoints for each control model are given in the Appendix. 
% ============================ MODEL SETUP ========================
\subsection{Model Setup and Evaluation}
The case study compares three future inverter deployment scenarios with either coordinated, autonomous or non-exporting inverters. The base case with 30\% PV penetration contains an equal share of autonomous and legacy inverters. Since the likelihood of experiencing voltage issues is location-dependent~\cite{Chou2017}, we randomly select 40 different PV locations. This set includes an instance with a PV cluster with the minimum electric distance from the secondary transformer and another with a cluster at the maximum distance. Then, we increase PV penetration in incremental steps of 10\% by adding one of three inverter types. The locations of new PV systems evolve from the base case and are consistent across inverter scenarios.

In this study, we consider that a VPP aggregator is in charge of coordinated inverter control and has access to the AMI data. The foremost task of the VPP aggregator is congestion management and the provision of grid support services within the network physical limits. A VPP aggregator is responsible for operating the assets in the most efficient manner while avoiding unnecessary power curtailment. Thus, we evaluate the effectiveness of the proposed coordinated inverter deployment in terms of voltage disturbances and energy losses. The performance metrics for voltage regulation are
\begin{itemize}
    \item number of customers experiencing overvoltages;
    \item total duration of overvoltages;
    \item maximum voltage observed for any customer;
    \item maximum voltage difference between consecutive time intervals for any customer;
    \item maximum voltage difference for each scenario. 
\end{itemize} 
and the performance metrics for energy optimisation are
\begin{itemize}
    \item PV output utilisation;  
    \item number of disconnection cases per inverter per type;
    \item total line losses;
    \item reactive power demand at the network-head. 
\end{itemize} 
Recall that the purpose of the paper is to demonstrate the extent to which coordinated inverter control can be used to maintain the voltage across the network below $\textbf{V}^\text{trip}$, which would trigger inverter tripping under the current reference settings in the Australian context. A model with lower a $\textbf{V}^\text{CIC}$ could provide stricter voltage bounds or enable grid support services as explained below.  
%
% ============================= GRID SUPPORT ========================
%
\subsection{Provision of Grid Support Services}
Coordinated inverters with bidirectional communication capability can be used for tasks beyond local voltage control. In particular, they could allow the VPP aggregator to respond to energy market price signals or provide ancillary services. However, there is a problem that is seldom mentioned in this context. At times of voltage stress, any change in the coordinated inverters' output will induce autonomous inverters to change their setpoints as a result of voltage change across the network. If an autonomous inverter was operating in Volt/Watt response mode before the market signal, then the voltage drop induced by coordinated inverters will allow autonomous inverters to ramp up power output. In this case, autonomous inverters will partially offset the down-regulation delivery by the VPP aggregator. This happens when the distribution network is operating close to the upper voltage limits, but with increasing PV penetration such events will occur for longer periods. Moreover, the fact that the response rate depends on the state of the distribution network adds to the challenge.

To address this, we examine the extent of such a power trade-off between coordinated and autonomous inverters following a request for grid support services (GSS) from the system operator. A VPP aggregator operates as a non-scheduled, price-taking market participant with grid support capability. This status allows the VPP aggregator to operate at any capacity level with the maximum up-regulation $\mathit{UR}$ and down-regulation $\mathit{DR}$ capacity \emph{offer}, determined by the VPP feasible operating region. The aggregator's decision for the capacity offer and direction would depend on the price forecast for GSS and local network constraints. This study excludes the monetary aspect of grid support services; instead it evaluates the trade-off between the system operator's up-regulation $\mathbf{UR}$ or down-regulation $\mathbf{DR}$ capacity \emph{request} and the response from autonomous inverters. The limits of the offered capacity and market response are explained next.
% ============================= GRID SUPPORT =============================
\subsubsection{Up- and down-regulation capacity}
%=========================================================================
The provision of the $\mathit{UR}$ service with coordinated inverters implies that inverters operate below their maximum output $\textbf{P}^\text{av}_c$. Moreover, a request to increase the output may also lift the voltages outside the operational limits in a PV-rich distribution network requiring further output reduction. We address that as follows. First, the VPP aggregator finds optimal inverter setpoints ($P^\text{inj}_c,Q_c$) from \eqref{eq:Objmain} that yield the upper limit of injected power $\sum_{c\in\mathcal{C}}P^\text{inj}_{c,t}$ and guarantee network operation without local overvoltages. In order to reserve the capacity for an $\textbf{UR}$ request, it next reduces the total power output by a factor $\gamma$ and calculates the updated setpoints for the new output limit.  Finally, these setpoints are used in the inverters themselves. This paper uses $\gamma = 0.2$, so that the VPP aggregator operates PV inverters at 80\% of the maximum feasible output. The choice of $\gamma$ in practice depends on the relative prices of energy and regulation. The maximum $\mathit{UR}$ and $\mathit{DR}$ offers are limited by 
\begin{equation}
    \mathit{UR}_t \leq \gamma\sum_{c\in\mathcal{C}}P^\text{inj}_{c,t},
    \label{eq:UR_offer}
\end{equation}
\begin{equation}
   \mathit{DR}_t \leq (1-\gamma)\sum_{c\in\mathcal{C}}P^\text{inj}_{c,t}. 
    \label{eq:DR_offer}
\end{equation}
In practice, advertised amounts would be less than these limits to account for the errors between forecast and expected PV output and household load values.

This section will quantify the response of autonomous inverters following the changes in coordinated inverter operation. A market designer would have to decide whether it was the responsibility of the VPP to control the inverters in such a way that the aggregate change matches the operator's request, or the operator's responsibility to request enough adjustment to account for the autonomous inverters.  This is beyond the scope of this study, although it is worth noting that the VPP is in a better position to estimate the response of the autonomous inverters.

% ============================= GRID SUPPORT
\subsubsection{GSS request}
%==============================
When the system operator issues a request at time $\tau$ to increase or decrease the coordinated inverter output within the offered capacity range, the VPP must comply and maintain it for 5 minutes through PV curtailment.  At each $t$ in this interval, the VPP aggregator will find new setpoints ($P^*_c,Q^*_c$) by solving an optimisation problem based on~\eqref{eq:Objmain} with the following additions. The regulation is enforced by one of the constraints
\begin{align}
    \sum_{c\in\mathcal{C}} \left(\textbf{P}^{\text{av}}_{c,t}-P^*_{c,t}\right)\geq (1-\gamma)\sum_{c\in\mathcal{C}} P^{\text{inj}}_{c,\tau-1}+\mathbf{UR}_\tau,% \quad \forall c \in \mathcal{C}, 
    \label{eq:verifyUR}
    \\
    \sum_{c\in\mathcal{C}} \left(\textbf{P}^{\text{av}}_{c,t}-P^*_{c,t}\right) \leq (1-\gamma)\sum_{c\in\mathcal{C}}P^{\text{inj}}_{c,\tau-1}-\mathbf{DR}_\tau,% \quad \forall c \in \mathcal{C},     
    \label{eq:verifyDR}
\end{align}
which ensure that in each dispatch time interval $t\in[\tau,\tau+5\text{ minutes}]$ the total injected output is different by at least $\textbf{UR}$ or $\textbf{DR}$ from the period before the GSS request, as appropriate.

Additional decision variables $P^\text{GSS}_{c,t}$ for $c\in \mathcal{C}$ satisfying
\begin{align}
    \textbf{UR}_t= \sum_{c\in\mathcal{C}} P^\text{GSS}_{c,t}, 
    \label{eq:UR_request} 
    \\
    -\textbf{DR}_t= \sum_{c\in\mathcal{C}} P^\text{GSS}_{c,t},
    \label{eq:DR_request}
\end{align}
and corresponding curtailment constraints
\begin{equation}
    \textbf{0} \leq  P^*_{c,t} + P^\text{GSS}_{c,t} \leq \textbf{P}^\text{av}_{c,t}, \quad \forall c \in \mathcal{C}
    \label{eq:FCAS_const_Pinj}
\end{equation}
are introduced to determine the optimal curtailment from each individual inverter when a request for $\mathbf{UR}$ or $\mathbf{DR}$ is issued. 

To demonstrate the network effects of the GSS delivery and the interaction between different inverter types, we simulate 320 market requests at 13 different PV generation levels, incrementally increasing the output from 40\% to full capacity. Based on the results in the first part of the study, we selected 60\% penetration level with 1:1 ratio between coordinated and autonomous. At this penetration, solar PV yields excess output between 20-120\;kW, sufficiently increasing the voltages to activate autonomous inverter response modes. 

The new decision variable $P^\text{GSS}_{c,t}$ should also be introduced in \eqref{eq:CICconst2}, \eqref{eq:NodalConst1}, and~\eqref{eq:NodalConst2} such that $P^\text{inj}_{c,t}$ is replaced by $P^\text{inj}_{c,t}-P^\text{GSS}_{c,t}$. Following the deployment of GSS across the coordinated inverters, the total change in the autonomous inverter output $\sum_{a\in\mathcal{A}} \Delta P^\text{inj}_{a,t}$ is recorded to quantify the actual curtailment as seen by the system operator. The results provide bounds on the actual curtailment that the VPP aggregator can offer:
\begin{align}
        UR_t \leq \gamma\sum_{c\in\mathcal{C}}P^\text{inj}_{c,t}- %\Tilde{P}^{\text{inj}}_{c,t}, 
        \sum_{a\in\mathcal{A}} \Delta P^\text{inj}_{a,t}, \\
       DR_t \leq (1-\gamma)\sum_{c\in\mathcal{C}}P^\text{inj}_{c,t} - \sum_{a\in\mathcal{A}} \Delta P^\text{inj}_{a,t}.
    \label{eq:FCAS_AUTresp}
\end{align}
 
% =========================================
\section{Results}\label{Results}
\subsection{Inverter Deployment Scenarios}
The primary objective for deploying coordinated inverter control (CIC) is to exploit its effectiveness in regulating voltage. Therefore, we first analyze voltage metrics observed in the test networks for three future inverter deployment scenarios: additional autonomous inverters~(AUT), additional non-exporting inverters~(Non-Exp), and additional coordinated inverters~(CIC). 

\begin{figure}[t]
    \centering
    \includegraphics[width=\linewidth]{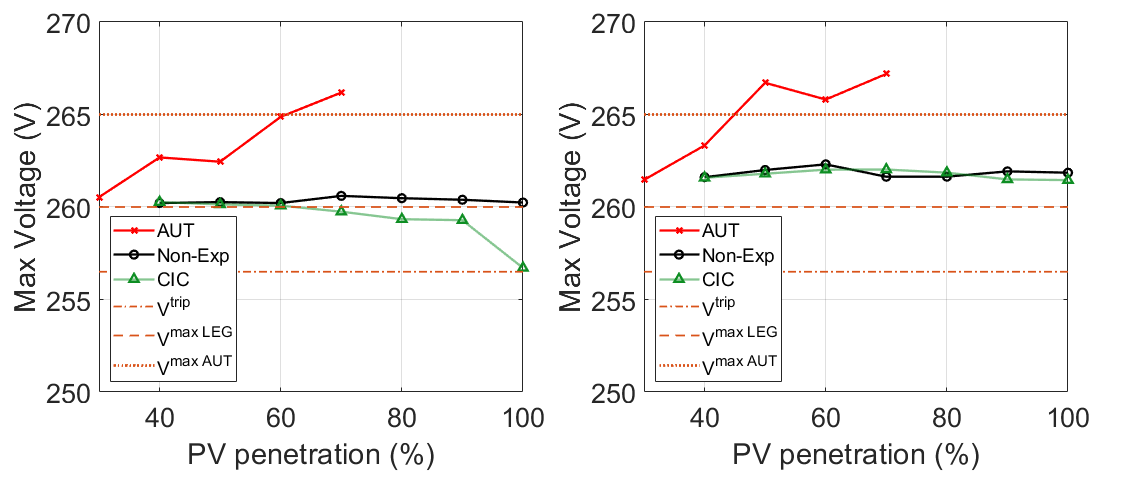}
    \caption{The maximum observed voltage on the 114-bus network (left) and 906-node network (right) observed for all PV distribution cases.} 
    \label{fig:maxV}
\end{figure}

Figure~\ref{fig:maxV} illustrates the highest recorded voltage for each network across all PV locations. The deployment of additional autonomous inverters leads to power system instability above 70\% penetration due to voltages exceeding the average tripping threshold. A larger time-constant $\tau$ value could potentially mitigate voltage oscillations, but this approach would also increase the delay in the inverters' response and violate droop curve specifications. Since autonomous inverters do not return to predisturbance power injection levels, the results for autonomous inverter-based penetration levels above 70\% are excluded. 
% === FIGURE Max V and Delta V
%maximum voltage difference between consecutive time intervals

Non-exporting inverters maintain the operation around the same level for all penetration levels above the base case. Limited to self-consumption, those inverters do not contribute to voltage fluctuations or overvoltages since each new non-exporting inverter makes the customer 'invisible' to the network operator (zero net load) at times of high PV generation. The difference in maximum voltage levels between the two networks is due to the higher line impedances of the 906-node network.

Coordinated inverter control demonstrates a downward trend in the maximum observed voltages for the 114-node network, approaching the CIC voltage threshold. At 100\% PV penetration, CIC is successful in preventing passive inverter tripping in 39 out of 40 PV locations, ultimately proving its capability to prevent overvoltages. However, PV clustering at the far end of the distribution network may limit the potential of CIC to mitigate voltage disturbances. In such a case, the VPP aggregator could gradually reduce the CIC limit and drive it towards the desired 230\;V +10\%. The high tripping threshold of legacy inverters is discouraging the VPP aggregator to set a lower level, thus curtailing extra power. However, the reasons for operating coordinated inverters at a lower voltage and maintaining reserve capacity for grid support services are demonstrated later in this paper. An alternative could be retrofitting of some existing legacy inverters to provide sufficient voltage control. This trend is likely to be seen in the upcoming years as the expected lifetime of legacy inverters is coming to an end.

A different outcome is observed on the 906-node network, where CIC maintains the maximum voltage at the base case level. This demonstrates that for the given network topology and loads, the base case penetration level of 30\% is too high for CIC to reverse voltage disturbances. In other words, it is too late to deploy CIC alone for voltage regulation on this network.

\subsubsection{Extent of effects}
\begin{figure}[t]
    \centering{\includegraphics[width=\linewidth]{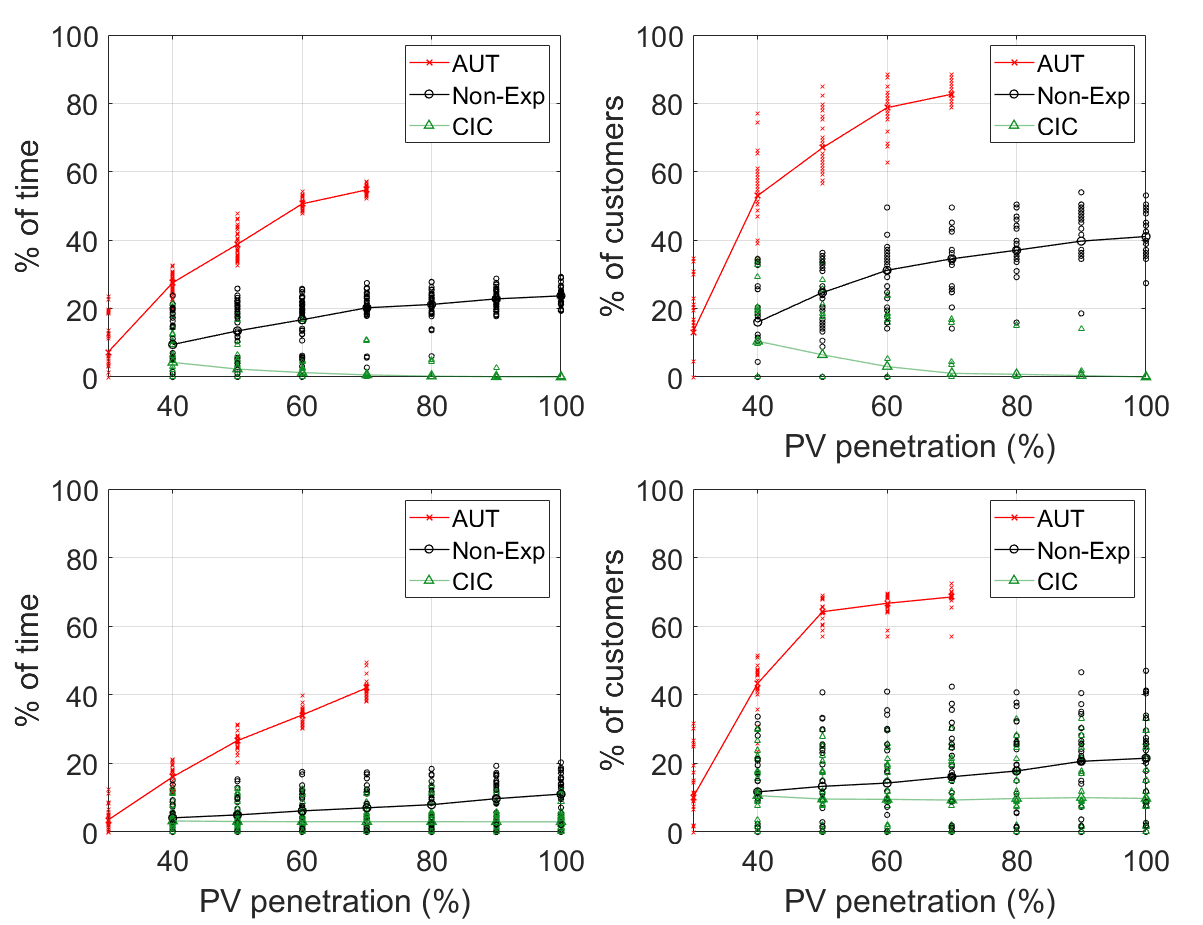}}
    \caption{The change in the length of time and the number of customers affected at increasing penetration with three types of inverters: autonomous, non-exporting and coordinated. The 114-node network (top) and 906-node network (bottom).} 
    \label{fig:Vcust_Vtime} 
\end{figure}

%We have seen the benefits of CIC in the worst-case performance; now let us consider performance away from this worst case. 
More insights into overvoltages can be obtained from Fig.~\ref{fig:Vcust_Vtime}. With the continuous deployment of autonomous inverters, overvoltages occur earlier in the day than with other types of inverters, affecting more than 60\% customers at 50\% PV penetration. Meanwhile, the deployment of non-exporting inverters does not increase reverse power flow, but more excess power from the existing inverters is exported to the medium voltage level as new PV installations with Non-Exp meet households' own demand. Thus, the number of affected customers at 50\% penetration for the 114-node and 906-node networks is 10\% and 5\% higher than in the base case, respectively. %With the deployment of autonomous inverters, the number of affected customers grows rapidly, while an opposite trend can be observed with coordinated inverters. CIC manages to reduce the duration of overvoltages on a day with high solar radiation to 0\% in 28/40 cases at 50\% penetration. 

Building on the previous observations, it can be seen that increasing the number of coordinated inverters reduces the number of customers and the length of time with voltage disturbances on the 114-node network while maintaining the same level of adverse network effects on the 906-node network. In terms of the minimum share of coordinated inverters required to fully arrest exceeding voltages induced by passive inverters, we observed that the optimal number depends on the base case penetration level at which the CIC is added and the PV locations across the network. Given a one-to-one ratio with passive inverters at 60\% penetration, CIC is effective to eliminate voltage disturbances for 85\% of PV locations on the 114-node network, while only 37\% of PV locations remain unaffected on the 906-node network. %At 100\% penetration, CIC is capable to prevent overvoltages in 39/40 cases; the exception is an instance in which multiple legacy inverters are clustered at the end of the line limiting the CIC's ability to mitigate overvoltages (with a mitigation strategy described earlier).
% ==================================
\subsubsection{Voltage oscillations}
Another performance problem caused by autonomous inverters is large, rapid changes in voltage due to inverters tripping and then resuming operation. Inverter cycling is mainly caused by the lack of advanced voltage control in legacy inverters. This results in both legacy and autonomous inverters exceeding the average tripping voltage $\textbf{V}^\text{trip}$ and occasionally by reaching the voltage $\textbf{V}^{\max}$ that necessitates instantaneous tripping. %The rapid increase in cycling cases between 30\% and 40\% PV penetration can be understood by looking at the network topology. At this level of installed PV capacity, overvoltages also start to occur on the branch ending with node-50 (Fig.~\ref{fig:network}), while at lower penetration customers are mostly affected along the branch starting at node-51.
Inverter cycling also increases with non-exporting inverters on both networks. Interestingly, these inverters will be occasionally shut off although not contributing to overvoltages. The frequent occurrence of such events could encourage customers to install stand-alone inverters and energy storage systems that would allow them to disconnect from the grid on such occasions. 

\begin{figure}[t]
    \centering{\includegraphics[width=\linewidth]{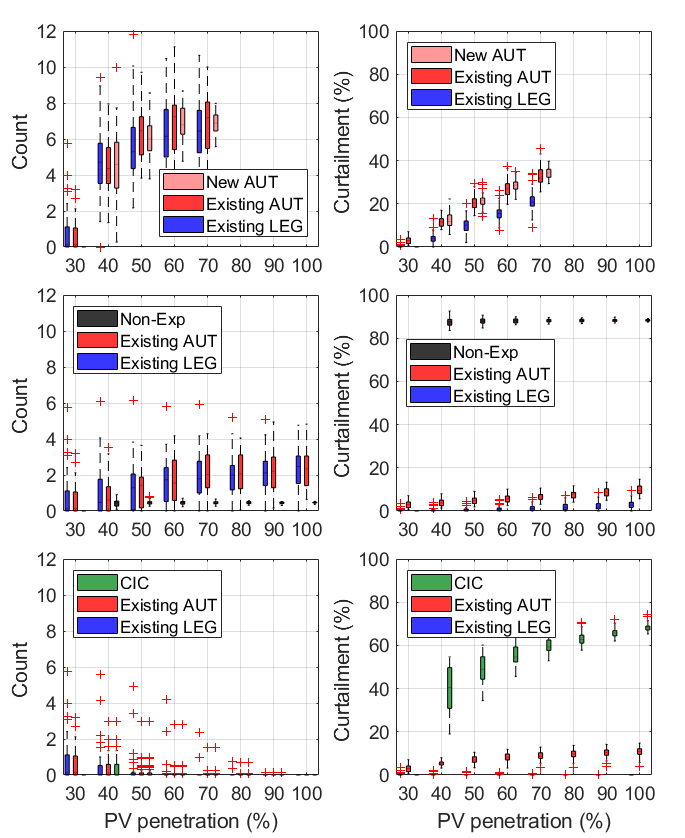}}
    \caption{The total number of disconnection cases per inverter and PV curtailment for autonomous (top), non-exporting inverters (middle), and coordinated (bottom) inverters. The 114-node network.} 
    \label{fig:invCurtCycl}
\end{figure}

However, different cycling patterns were observed with CIC on each network. On the 114-node network, CIC reduced cycling significantly or even eliminated it for both the existing and new PV installations (Fig.~\ref{fig:invCurtCycl}). As mentioned above, CIC cannot completely eliminate cycling of legacy inverters; when CIC inverters are curtailing 100\% and drawing the maximum reactive power, they cannot do more to prevent legacy inverters from tripping. Higher line impedances in the 906-node network leads to more cycling cases at the same penetration level simply due to an insufficient number of coordinated inverters.  
\subsubsection{PV output utilisation}
In this section, we show that PV curtailment is not necessarily a valid metric to assess the effectiveness of coordinated inverters. The goal of the VPP aggregator is to maintain the network within operational limits while maximising the revenue from the available solar generation. The utilised PV output is perceived as the sum of the PV output consumed locally and exported via the distribution transformer. During periods when the network is not constrained, the optimal setpoints for coordinated inverters are determined by minimising active power losses. Adding a term for line losses to the cost function is equivalent to maximising the utilised PV output. 

\begin{figure}
    \centering{\includegraphics[width=\linewidth]{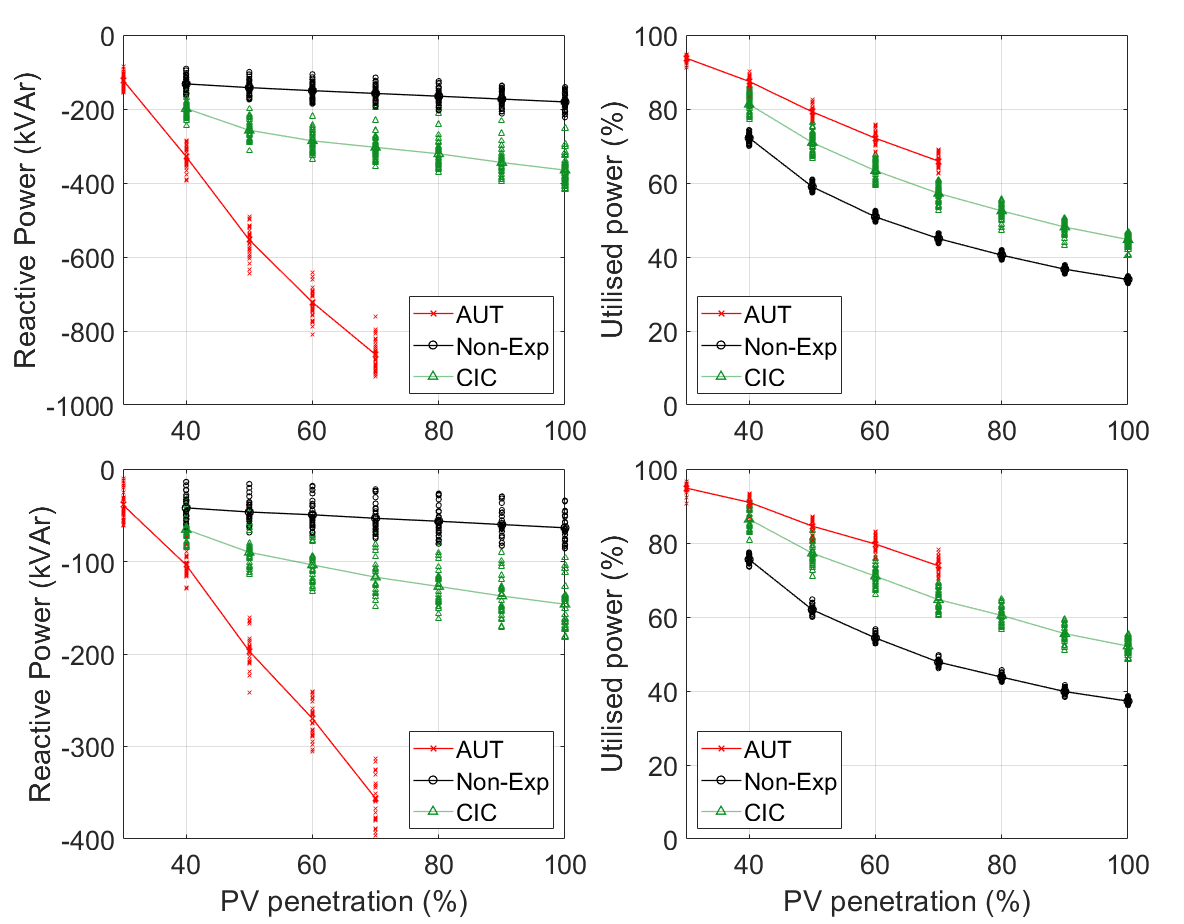}}
    \caption{Total reactive power demand (left) and the utilised PV output (right) for the 114-node network (top) and the 906-node network (bottom).} 
    \label{fig:Qd_Putil}
\end{figure}

For example, the higher reactive power demand from autonomous inverters operating in Volt/VAr or Volt/Watt mode at the far end of the line may increase line power losses more than the higher PV output as a result of lower voltages. Additional line losses cause increased heating of the conductors, which increases the sag; this may present a fire risk on hot days if there are trees under the power lines. Moreover, the increased apparent power increases the thermal stress on the transformers, which are already stressed on summer days. Since the CIC objective is to minimise total power losses, reactive power from coordinated inverters is used only when it leads to lower objective values (Fig.~\ref{fig:Qd_Putil}).
% === FIGURE ===
Figure~\ref{fig:invCurtCycl} illustrates the difference in curtailment for each of the future scenarios. If the VPP aggregator considered only PV output curtailment, the future scenario with coordinated inverters would lead to a 20-30\% higher curtailment. However, when looking at PV output utilisation, this gap between the same scenarios is reduced to approximately 10\% due to CIC yielding lower line losses (Fig.~\ref{fig:Qd_Putil}). In addition, higher PV utilisation by autonomous inverters is achieved only at the cost of voltage violations. Distribution network service providers (DNSPs) are likely to prevent this from happening by limiting the number of PV connections in the network, and may also consider network augmentation or the deployment of dedicated voltage regulation hardware as a potential solution. Coordinated inverter control can achieve similar benefits while utilising the existing infrastructure. The additional curtailment, however, highlights the need for a mechanism that allows the VPP aggregator to partially compensate adversely-affected PV owners for providing local voltage regulation. 

Meanwhile, non-exporting inverters may curtail up to 90\% of the available generation on a given day. To mitigate these effects, PV owners may try to increase their loads, invest in energy storage or opt for a smaller PV system to minimise the wasted PV output. As adding more non-exporting inverters reduce customer loads, the PV output exports from autonomous and legacy inverters results in a higher reverse current flow, thus increasing passive inverter cycling. 

\begin{figure}[H]
    \centering
     \includegraphics[width=\linewidth]{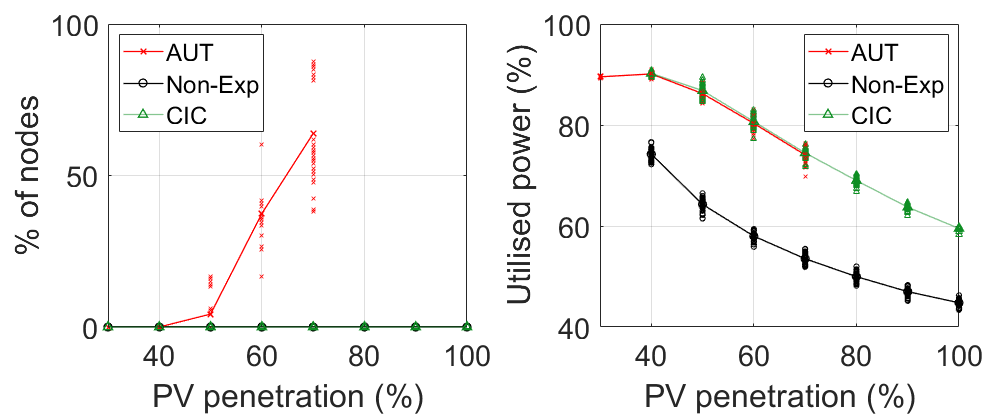}
    \caption{The number of customers with overvoltages (left) and the utilised power as the percentage of total solar PV generation (right) from all types of inverters on a clear winter day. The 114-node network.} 
    \label{fig:winter}
\end{figure}

The results demonstrated so far represent a summer day with high PV generation and low load. Figure.~\ref{fig:winter} illustrates inverter operation on a winter day with approximately 30\% less solar irradiation compared to summer. CIC is capable of maintaining the voltage below the target at all penetration levels, while autonomous inverters lift the voltage above the desired operating limits when exceeding 40\% penetration. In the absence of voltage disturbances, CIC offers marginal gains in utilised power over autonomous inverters due to better line-loss and reactive power management (Fig.~\ref{fig:winter}). As there are more winter-like days in the year (in terms of available solar irradiation), CIC has shown to yield as much power as autonomous inverters in those periods, while preventing voltage issues during the days with high PV output. 
% === Winter
% ============================= SUBSECTION ====================
\subsection{Grid Support Services}
\begin{figure}[tb] 
\centerline{\includegraphics[width=\linewidth]{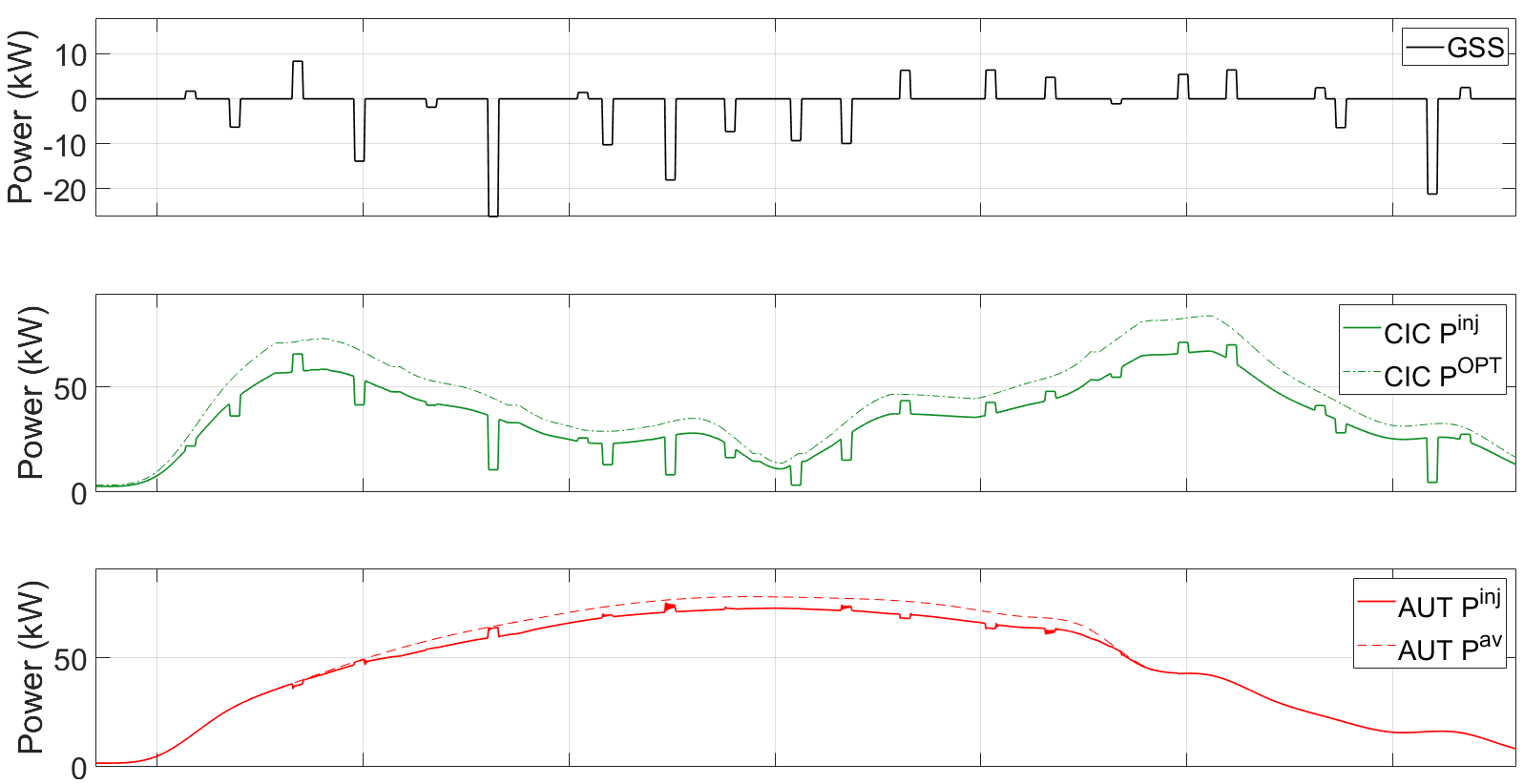}}
\caption{The change in active power output from autonomous and coordinated inverters following a GSS request.} 
\label{fig:DR_Presponse} 
\end{figure}

% ============================= SUBSECTION ====================   
% ============================= Down-regulation ====================
To study the use of coordinated inverters for grid support services (GSS) we populate the 114-node LV distribution network with a number of coordinated inverters and passive inverters replicating one of the voltage disturbance-free PV layouts at 60\% penetration. The provision of GSS is illustrated in Fig.~\ref{fig:DR_Presponse} with $\mathbf{UR}$ and $\mathbf{DR}$ requests received at random time intervals over the daylight hours. After receiving a GSS request (Fig.~\ref{fig:DR_Presponse}, top), coordinated inverters respond to it by adjusting the power output (Fig.~\ref{fig:DR_Presponse}, middle). The VPP aggregator's capability to provide GSS during peak sun-hours is significantly reduced due to the additional curtailment required for keeping voltages below the tripping threshold. The deployment of GSS leads to voltage drop/rise along the distribution lines causing autonomous inverters to increase/decrease their power injection according to droop curve settings (Fig.~\ref{fig:DR_Presponse}, bottom). 
% === FIGURE ===
% === FIGURE ===
\begin{figure}[b] 
\centerline{\includegraphics[width=6cm]{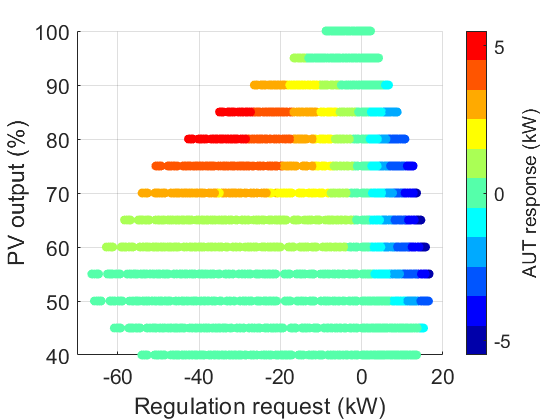}}
\caption{The response rates of autonomous inverters at various generation to load ratios, following a GSS request.} 
\label{fig:GSS} 
\end{figure}
% === FIGURE ===
\begin{figure}[b] 
\centerline{\includegraphics[width=6cm]{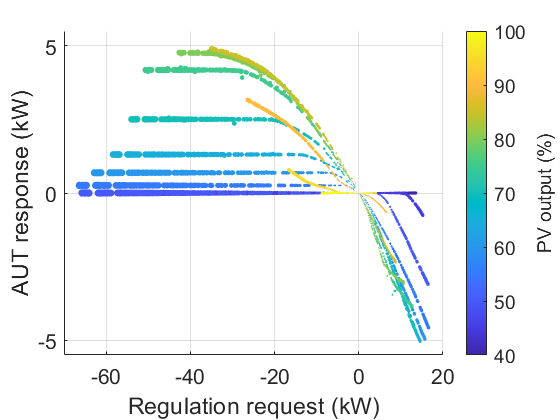}}
\caption{The response rates of autonomous inverters at various generation to load ratios, following a GSS request.} 
\label{fig:GSS2} 
\end{figure}
% === FIGURE ===
\begin{figure}[b] 
\centerline{\includegraphics[width= 6cm]{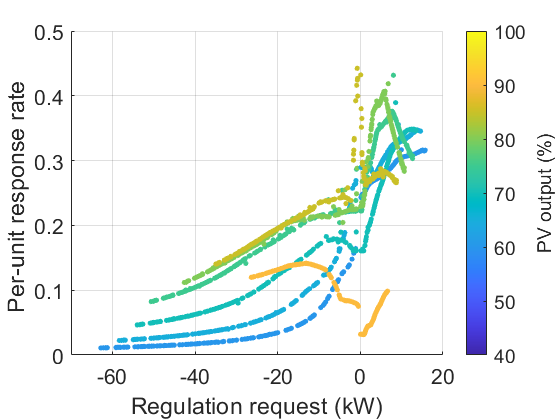}}
\caption{The autonomous inverter response rate per unit of the regulation request value between 50-90\% PV output.} 
\label{fig:GSSperunit} 
\end{figure}
% =================
\subsubsection{Passive inverter response}
% =================
The operation of legacy inverters is not affected by the VPP aggregator providing GSS. The VPP aggregator prioritises local voltage regulation through own curtailment optimisation in order to maintain all PV resources online. Since the maximum $\mathit{UR}$ offer capacity is calculated for the optimum operation of coordinated inverters at full output, the delivery of any $\mathit{UR}$ request will keep voltages below the inverter disconnection threshold allowing legacy inverters to operate at unity power factor. Therefore, a legacy inverter would only shut off once all coordinated inverters have reduced power injection to zero and have disconnected from the grid. 

% =================
\subsubsection{Autonomous inverter response}
% =================
The autonomous inverters will adjust operating points according to the voltage difference prior to and after a GSS event. At each time interval, the initial voltage level is driven by the amount of excess solar output that is being pushed back to the transformer, whereas the GSS dispatch capacity will determine the level at which voltage settles. Combining these two factors, the autonomous inverter response is illustrated in Fig.~\ref{fig:GSS}. PV output levels are varied between 40-100\% with constant customer load. The total width of each $\mathbf{UR}$ and $\mathbf{DR}$ band corresponds to the VPP feasible operating region (FOR) at the given PV output. The size of FOR follows the increase in solar generation, peaking at 55\%. Above this level, the VPP aggregator is required to curtail additional power from coordinated inverters (can be observed in Fig.~\ref{fig:DR_Presponse}) in order to maintain voltages below the disconnection threshold across the network. 

% ================================
The $\mathbf{DR}$ delivery affects autonomous inverters when the PV output is between 60-90\%. Voltage drop induced by coordinated inverters allows autonomous inverters to either move up the Volt/Watt generation curve or deactivates them completely. Autonomous inverters recover up to 5\;kW when a $\mathbf{DR}$ request is between 30-45\;kW. It is important to note that the total response rate of autonomous inverters is limited by the curtailment occurring before a GSS event. Figure~\ref{fig:GSS2} illustrates the response rate against the regulation value for each PV output level. Depending on the PV output level, autonomous inverter response becomes constant for increasing $\mathbf{DR}$ values when inverters have recovered to full output. Then, the flat segment of each curve sets the upper bound of the $\mathbf{DR}$ response rate.

The impact of the $\mathbf{UR}$ request on autonomous inverters stretches across a wider range of PV output than during $\mathbf{DR}$. By delivering the $\mathbf{UR}$ request, CIC lifts the initial voltages to within the Volt/Watt response region. This causes autonomous inverter setpoints to move away from full output or increases the pre-existing curtailment. The total response rate from autonomous inverters increases together with $\mathbf{UR}$ request capacity. Since the voltage range of the Volt/Watt response mode exceeds the CIC tripping threshold, a larger $\mathbf{UR}$ request than the available reserve capacity would further reduce the output from autonomous inverters. However, it would also compromise the ability of CIC to maintain voltages below the average tripping level.  

It may be argued whether the idea of providing up-regulation from PV inverters is reasonable. Operating coordinated inverters below the optimal output in the expectation of an $\mathbf{UR}$ request leads to unnecessary energy curtailment. However, without an option to consume or store excess PV output locally, it may have low or no market value anyway. The energy in wholesale electricity and ancillary markets is frequently traded at zero price during peak-sun hours. With the number of solar PV installations far exceeding the uptake of battery storage for a foreseeable future, the price received for $\mathbf{UR}$ is likely to be greater than for the $\mathbf{DR}$ capacity. The fact that operating coordinated inverters below the optimum level may relax the voltage constraints also supports the idea of operating coordinated inverters with reserve capacity. Further research should focus on introducing a dynamic ratio between the $\mathit{UR}$ and $\mathit{DR}$ offer. This would enable the VPP aggregator to maximise the revenue while operating coordinated inverters and DER within the network limits. 

The autonomous inverter response rate per unit of GSS request is illustrated in Fig.~\ref{fig:GSSperunit}. For each kW of the $\mathbf{DR}$ request, autonomous inverters may increase the output by up to 0.25\;kW when operating at 85\% of PV output. The response rate increases up to 0.45\;kW during up-regulation. The difference between $\mathbf{UR}$ and $\mathbf{DR}$ response levels is related to the response being bounded by the total autonomous inverter curtailment and near-zero regulation requests. Although occurring at specific PV output levels, the extent of autonomous inverter response is large enough to derail the GSS delivery. The VPP aggregator, likely the only entity with controllable assets in the distribution network, should consider this counter-effect when operating in a network with a significant share of autonomous inverters. As shown before, at higher penetration levels, distribution networks will operate within the Volt/Watt mode more frequently and for longer time periods. If the actual load/output change seen by the system operator is only 55-75\% of the requested capacity, this undermines the utilisation of DER for grid support services.

 \section{Conclusions}\label{Conclusions}
This paper studied the operation of coordinated inverter control deployed in two different low-voltage test networks with an existing 30\% PV penetration comprised of passive inverters. The number of customers experiencing overvoltages and the duration of those voltage disturbances were reduced as more coordinated inverters were added to the central control model on the 114-node network. At a one-to-one ratio between coordinated and passive inverters, voltage issues were mitigated at 34 of the 40 PV locations. This increased to 39 of the 40 locations at 100\% penetration. However, at the same penetration level, no voltage disturbances were observed at only 15 out of 40 PV locations on the 906-node network. This suggests that the studied networks have a penetration level beyond which coordinated inverter deployment cannot reverse voltage disturbances. It will be interesting to study how to determine this penetration threshold. 

Considering a one-to-one ratio between coordinated and passive inverters on the 114-node network, we examined an option to use the aggregated output from coordinated inverters as a Virtual Power Plant and provide grid support services. It was shown that the output change from autonomous inverters following a down-regulation request may increase by up to 0.2 units and decrease following an up-regulation request by up to 0.45 units at certain PV output levels. Since the regulation request and the response from autonomous inverters are of opposite sign, this difference has to be accounted for by the VPP aggregator. In future work, we aim to integrate the calculation of response rate into the VPP operation. %A dynamic ratio between up- and down-regulation following the price forecast should be also examined. 

% =============================FORMATTING GUIDELINES BELOW =================================

% \section{Acknowledgments}

% Identification of funding sources and other support, and thanks to
% individuals and groups that assisted in the research and the
% preparation of the work should be included in an acknowledgment
% section, which is placed just before the reference section in your
% document.

% This section has a special environment:
% \begin{verbatim}
%   \begin{acks}
%   ...
%   \end{acks}
% \end{verbatim}
% so that the information contained therein can be more easily collected
% during the article metadata extraction phase, and to ensure
% consistency in the spelling of the section heading.

% Authors should not prepare this section as a numbered or unnumbered {\verb|\section|}; please use the ``{\verb|acks|}'' environment.

\begin{acks}
This work was supported by ARC grant DP190102134.
\end{acks}

\bibliographystyle{ACM-Reference-Format}
\bibliography{Coordinated.bib}

\appendix
\section*{Appendix}
% ====== TABLE ==========
Table~\ref{tab:Setpoints} provides reference setpoints for legacy inverters and autonomous inverter Volt/VAr and Volt/Watt droop curves. The time constants $\tau^\text{V}$ and $\tau^\text{W}$ are set to 1.5 and 3.5, respectively. 
\begin{table}[H]
\caption{Inverter active and reactive power setpoints and voltage reference values.}
\label{tab:Setpoints}
\centering
\begin{tabular}{l l l l l l}
\toprule
Ref. & $V$ & $\textbf{Q}^{\min,pu}_a$  & $\textbf{P}^{pu}_a$ & $\textbf{P}^{pu}_l$
& $\textbf{P}^{pu}_c$ \\
\midrule
$\textbf{V}^{\min}$ &207& n/a & n/a & n/a & n/a\\
$\textbf{V}^\text{nom}$ &230& n/a & n/a & n/a & n/a\\
$\textbf{V}^\text{DB}$ &248& 0 & 1.0 & - & n/a\\
$\textbf{V}^\text{Qmin}$ &253& 0.44 & 1.0 & - & n/a\\
$\textbf{V}^{\max}_l$ &260& - & - & 0 & n/a \\
$\textbf{V}^{\max}_a$ &265& 0.44 & 0.2 & - & n/a \\
\midrule
$\textbf{V}^\text{CIC}$ &255.85& - & - & - & 0  \\
$\textbf{V}^\text{trip}$ &257& 0 & 0 & 0 & - \\
\bottomrule\\
\end{tabular}
\end{table}
%
% \begin{table}[h!]
% \caption{Active power $P^\text{PU}$ and Reactive power $Q^\text{PU}$ setpoints in per-unit.}
% \label{tab:active}
% \centering
% \begin{tabular}{l l l l l}
% \toprule
% \tabhead{Ref.} & \tabhead{$V$} & \tabhead{$Q^\text{PU}_a$}  & \tabhead{$P^\text{PU}_a$} &\tabhead{$P^\text{PU}_l$} \\
% \midrule
% $\textbf{V}^\text{nom}$ &230& 0 & 1.0 & 1.0 \\
% $\textbf{V}^\text{DB}$ &248& 0 & 1.0 & n/a \\
% $\textbf{V}^\text{Qmin}$ &253& 0.44 & 1.0 & n/a \\
% $\textbf{V}^\text{max}_l$ &260& n/a & n/a & 0  \\
% $\textbf{V}^\text{max}_a$ &265& 0.44 & 0.2 & n/a  \\
% \midrule
% $\textbf{V}^\text{trip}$ &257& 0 &  0 & 0  \\
% \bottomrule \\
% \end{tabular}
% \end{table}

\end{document}